\documentstyle[osa,manuscript,epsfig]{revtex}
\begin{document}

\title{Wide viewing angle realization for sampled hologram by collecting high-order diffraction beams}

\author{Byung Gyu Chae}

\address{Basic Research Laboratory, Electronics and Telecommunications Research Institute, 218 Gajeong-ro,
 Yuseong-gu, Daejeon 305-700, South Korea}

\maketitle{}
\begin{abstract}

We propose that
viewing angle expansion of the holographic image can be realized by using high-order diffraction beams
caused by the pixel structure sampling the hologram data.
The diffractive beam propagating to new optical axis direction plays a role in a modulated carrier similar to
a carrier signal of the off-axis holography,
which makes new viewing zone of the reconstruction image.
The reconstructed image in the Fresnel hologram is deformed along new viewing direction,
whereas the Fourier hologram enables to retrieve three-dimensional image with other perspective.
High-resolution hologram fringe is imaged on the image plane through an imaging system,
and thus, only collection of diffracted beams increases a viewing zone angle.
We verify our proposal through the numerical analysis for the sampled hologram
showing high-order diffraction beams with various viewing zones.


\end{abstract}

\section{Introduction}
\label{INT}

 The hologram implies three-dimensional information with phase and amplitude of the propagating wave [1,2].
This makes it possible to retrieve the wavefront of original propagating wave
by illuminating the hologram with reconstructing wave.
However, in digital holography,
the holography has difficulty in the recording and reconstructing processes of the hologram image
because of high spatial frequency of the fringe pattern [3-5].
The hologram is the interference fringe of the recording wave and the object wave,
and therefore, for retrieving the hologram image with sufficient viewing angle
the spatial frequency gets to the extent of the wavelength of the light field.

 In holographic display,
the hologram is sampled with sampling frequency of the device with pixel structure
to avoid aliasing effects [6,7].
Present modulator encoding hologram fringe has much lower resolution than the required specification.
Now, the limitation of the spatial frequency due to the device performance
obstructs widening view angle of the hologram image,
although the holographic display can reconstruct the image to the display size.
For the realization of holographic display, solving the problem of widening viewing angle is essential.

 Current researches for improving the viewing zone have been carried out
through temporal and spatial multiplexing of the spatial light modulators [8-10].
Multiplexing method of the hologram effectively enhances the viewing angle of the reconstructed image,
whereas the large space-bandwidth product is still required.
Since only the high-definition resolution of the real image is sufficient,
the use of the massive hologram data for generating holographic high-definition image is unreasonable.
This also hinders the process of the digitized hologram data.
Recently, the interesting technique by using resolution redistribution of the hologram fringe
was proposed to enlarge the viewing angle without a multiplexing process.
The duplicated hologram fringe with increased resolution can be imaged on the image plane through an imaging system,
which is implemented by a combination of the multiple point light sources and a rearrangement
of the hologram data.
Here, only the enhancement of the horizontal parallax is possible.

In this research, we note that in principle,
even the hologram with small space bandwidth has entire information of the object,
and thus trying to find how to enlarge the viewing zone of the reconstructed image optically.
First, for the Fresnel and Fourier holograms we investigate the possibility of the viewing angle change
by varying an incidence angle of the illuminating plane wave.
The plane wave illuminating the hologram in other direction than that of the recording process
generates diffracted beams emanating to the shifted position of the reconstructed image,
which can make the viewing zone with another perspective of the image.
Next, in the sampled hologram with pixel structure
the method of extending the viewing zone angle by collecting high-order diffracted beams is proposed.
Finally, we carry out the numerical analysis for the sampled hologram
showing high-order diffraction beams with various viewing zones.

\section{Viewing angle analysis of hologram image}
\subsection{Reconstruction process by a tilted plane wave}

 The reconstruction process of the objective wave in the on-axis hologram illuminating
by a tilted plane wave, $\exp(jk_{0}x \sin \theta)$
is expressed as the product of a hologram component, $a(x,y) \exp[j\phi (x,y)]$ and the illuminating wave.

\begin{equation}
a(x,y) \exp[j \phi (x,y)] \exp(j k_{0} x \sin \theta)
\end{equation}
When the plane wave with unit amplitude is used in recording the on-axis hologram,
$a(x,y)$ and $\phi(x,y)$ are real amplitude and phase of the object wave, respectively.
Equation (1) can be rewritten as another form.

\begin{equation}
a(x,y) \exp \left\{ j[\phi (x,y) + k_{0} x \sin \theta] \right\}
\end{equation}
This is a kind of the reconstruction form for the off-axis hologram made
by the recording plane wave, $\exp(jk_{0}x \sin \theta)$,
where the coaxial reconstruction wave is assumed.
Equation (2) can be also rewritten by varying together incidence angle values of the recording and
reconstructing plane waves in the off-axis hologram.
In a view of the carrier frequency holography,
the on-axis hologram fringe is unique
because it has a constant form as the modulating signal irrespective of the spatial carrier.
Therefore, in two cases of the on-axis holography by an inclined plane wave and the off-axis holography,
there is only the variation of the carrier frequency.

\begin{equation}
f_{c} = \frac{\sin \theta} {\lambda}
\end{equation}

 Figure 1 depicts the reconstruction by illuminating the on-axis hologram with the plane wave
at an incidence angle, $\theta$.
Here, we notice only the real image from the complex amplitude modulation
in order to clarify our analysis.
The viewing angle of the image by the coaxial plane wave is determined by means of the spatial frequency
of the fringe pattern.
Since the local spatial frequency is given by $\frac{x}{\lambda z_{0}} \approx \frac{\sin \omega}{\lambda}$,
the viewing angle, $\Omega$ is double the size of $\omega$.
The higher spatial frequency leads the wider viewing angle of the image.
In the tilted plane wave,
since the exponential term of a carrier frequency generates the deflected beam of the retrieved image,
the viewing zone of the object image changes to new $\Omega_{NEW}$ region in Fig. 1.

 We find that the tilted plane wave plays a role in a modulated carrier
similar to a carrier signal of the off-axis holography,
which makes the diffractive wave propagating to new optical axis direction.
In the sampled hologram,
the viewing zone of the reconstructed image has a limitation
because it has the bandlimit signal to the Nyquist frequency to avoid the aliasing effect [11].
But, the viewing zone variation of the object image by a tilted plane wave is irrespective of
the aliasing in principle.
This case is different from the aliasing arising in sampling the modulated carrier signal to the hologram.
Here, the modulating hologram signal of the object is already recorded in the sampled hologram,
and only the inclined reconstruction wave related to the carrier signal is illuminated.
Thus, by controlling the incidence angle of the reconstruction wave for the on-axis sampled hologram,
we can get the reconstruction effect of the off-axis holography making a new viewing zone.

\subsection{Hologram image deformation by a tilted plane wave}

 Let us investigate the change in the reconstructed image of the hologram transparency,
$g(x,y)$ by a tilted plane wave.
The complex field of the image in the Fresnel propagation is developed as the convolution of the input field,
$\exp[jk_{0} (x \sin \theta + y \sin \varphi)] g(x,y)$
and the spatial impulse response,
$h(x,y;z_{0}) = \frac{1}{j \lambda z_{0}} \exp(-j k_{0} z_{0}) \exp \left[-j \frac{k_{0}}{2z_{0}} (x^2 + y^2) \right]$.

\begin{eqnarray}
U(x,y;z_{0}) =  \frac{e^{j k_{0} z_{0}}}{j \lambda z_{0}} \exp \left[-j \frac{k_{0}}{2z_{0}} (x^2 + y^2) \right]
\int\!\!\!\int \exp[jk_{0} (x' \sin \theta + y' \sin \varphi)] g(x',y') \nonumber\\
\times \exp \left[-j \frac{k_{0}}{2z_{0}} (x'^2 + y'^2) \right]
\exp \left[j \frac{2\pi}{\lambda z_{0}} (xx' + yy') \right] dx' dy'
\end{eqnarray}
The integral part can be calculated from the convolution operation of two Fourier transforms
of the product term of the hologram and the plane wave and the quadratic phase term.

\begin{equation}
G \left[ \frac{1}{\lambda z_{0}} (x + z_{0} \sin \theta, y + z_{0} \sin \varphi) \right] \ast H(f_{X},f_{Y})
\end{equation}
where $H(f_{X},f_{Y})$ is the transfer function in the Fresnel diffraction.

\begin{equation}
\int\!\!\!\int G(f'_{X} + f_{Xc},f'_{Y} + f_{Yc}) H(f_{X} - f'_{X},f_{Y} - f'_{Y}) df'_{X} df'_{Y}
\end{equation}
By inserting Eq. (6) into Eq. (4),
above equation is the angular spectrum representation of the field propagation in the Fresnel diffraction.
If we substitute the variables, $f'_{X} + f_{Xc}$ by $f''_{X}$ and $f'_{Y} + f_{Yc}$ by $f''_{Y}$ and
rearrange the equation,
it becomes the expression for the complex field of the object shifted to the transversal axis.

\begin{equation}
U(x + z_{0} \sin \theta, y + z_{0} \sin \varphi ; z_{0})
\end{equation}

 The tilted plane wave causes the image reconstructed at the shifted position to the transversal axis.
This analysis is applicable to the arbitrary incidence angle of the tilted plane wave,
because the diffraction beam to retrieve an image can be regarded as the paraxial propagation
along the new diffraction optical-axis.
Figure 2(a) shows two-dimensional plane image vertical to optical axis retrieved at the shifted position
by an inclined plane wave to $x$-axis.
We find that in a plane object, new viewing angle of the image is generated,
because only the lateral position of the reconstructed image is shifted without distortion of the object shape.

 However, in a three-dimensional object, a change of an image shape occurs.
The three-dimensional object image can be calculated at various $z$ values.

\begin{equation}
\sum_{l} U(x + z_{l} \sin \theta, y + z_{l} \sin \varphi ; z_{l})
\end{equation}
The amount of shifted value to transversal axis depends on the $z$ value,
which invokes an image deformation along the depth direction.
The retrieved image of the hologram composed of the point sources in $x$-$z$ plane is illustrated in Fig. 2(b).
The point image is obtained from the simple convolution expression about one-point hologram [12].

\begin{eqnarray}
\exp(jk_{0}x \sin \theta) \exp \left[j \frac{k_{0}}{2z_{0}} (x^2 + y^2) \right]
 \ast h(x,y;z_{0}) \nonumber\\
\sim \int\!\!\!\int \exp \left\{j \frac{k_{0}}{z_{0}} \left[ (x+z_{0} \sin \theta)x'
  + yy' \right] \right\} dx' dy'
\sim \delta(x+z_{0} \sin \theta,y)
\end{eqnarray}
For convenience, the coefficient of each term shall drop.
The first line of above equation is also presented as the reconstruction of the off-axis hologram as like
$\exp \left\{ j \left[ -\frac{k_{0}}{2z_{0}} (x^2 + y^2) + k_{0}x \sin \theta \right] \right\}
\ast h(x,y;z_{0})$.
The real image appears as the delta function.
If $\sin \theta$ is written by $\frac{x_{p}}{z_{0}}$, the point image moves to $x_{p}$ amount at $x$ axis.
Four real images are reconstructed at following locations,
where the interval between two points in the axial direction is put to be $c$.

\begin{eqnarray}
\delta \left(x-\frac{c}{2}+x_{p1},y;z_{0}-\frac{c}{2} \right)
 + \delta \left(x+\frac{c}{2}+x_{p1},y;z_{0}-\frac{c}{2} \right) \nonumber\\
 + \delta \left(x-\frac{c}{2}+x_{p2},y;z_{0}+\frac{c}{2} \right)
 + \delta \left(x+\frac{c}{2}+x_{p2},y;z_{0}+\frac{c}{2} \right)
\end{eqnarray}
where $x_{p1}$ and $x_{p2}$ are the values, $(z_{0}-c/2)\sin \theta$ and $(z_{0}+c/2)\sin \theta$.

 As shown in Fig. 2(b),
shearing transformation arises along $x$-axis.
The image deformation by a tilted reconstruction plane wave,
$\exp[jk_{0} (x \sin \theta + y \sin \varphi)]$ is described in terms of Affine transformation.

\begin{equation}
\left(\begin{array}{c} x' \\ y' \end{array} \right)
 = \left(\begin{array}{cc} 1 & m \\ n & 1 \end{array} \right)
 \left(\begin{array}{c} x \\ y \end{array} \right)
 +\left(\begin{array}{c} z \sin \theta \\ z \sin \varphi \end{array} \right)
\end{equation}
We find that the reconstruction plane wave incident on at a different angle from
that of the recording process definitely induces the deformation of a retrieved image,
which holds for both on-axis and off-axis holograms.
Although the plane object image makes a new viewing zone in terms of an inclined plane wave,
three-dimensional image does not form new entire view in the Fresnel hologram.
That is, at a viewing direction observer sees the projection scene of reconstruction image
by coaxial plane wave with respect to vertical plane of deflection axis, but deformed image shape.

\subsection{Viewing angle change for the Fourier hologram image}

 Figure 3 illustrates the reconstruction process for the Fourier hologram by a tilted plane wave.
The hologram plane is placed in the front focal plane of the convex lens of focal length, $f$.
The field distribution of the image in the vicinity of the back focal plane, $d$ is obtained
from below equation [13,14].

\begin{eqnarray}
U(x,y;\Delta z) =  \frac{1}{j \lambda f}
\int\!\!\!\int \exp[jk_{0} (x' \sin \theta + y' \sin \varphi)] g(x',y') \nonumber\\
\times \exp \left(-j\frac{2 \pi}{\lambda} \left[ \frac{(x'^2 + y'^2) \Delta z}{2f^2}
 - \frac{(xx' + yy')}{f} \right] \right) dx' dy'
\end{eqnarray}
where $\Delta z$=$f$-$d$ indicates the image depth distribution.
From the similar interpretation of Eqs. (4)-(7) in the Fresnel hologram,
we can get the expression for the complex field of the object image shifted to the transversal axis.

\begin{equation}
U(x + f \sin \theta, y + f \sin \varphi ; \Delta z)
\end{equation}
Here, the shift quantity to the transversal axis has a constant value as $f \sin \theta$ or $f \sin \varphi$
irrespective of the image depth, and thus, as different from the Fresnel hologram,
the image deformation does not occur along the viewing direction.

 The tilted plane wave does not change the shape of the reconstructed image
and only moves the image to lateral position, as depicted in Fig. 3.
Therefore, the inclined plane wave can vary effectively the viewing angle of some object image
reconstructed optically as a function of the incidence angle in the Fourier hologram.
We clearly confirm that the perspective of the object image changes according to the propagation direction
of diffracted beams.
Considering the virtual image formation with opposite view of the real image,
it is apparent that even the hologram with extremely low spatial-frequency shall retrieve
the whole view of the object image optically in a transparent object.
This is possible because the hologram has entire information of three-dimensional object.

\section{Expansion of viewing angle in sampled Fourier hologram}
\label{EVS}

 The pixel structure of spatial light modulator sampling the hologram data generates high-order diffracted beams.
The diffracted beams propagate in the direction at an angle with respect to the optical $z$-axis.
We investigate the possibility of widening viewing angle by collecting high-order diffracted beams.
Figure 4 illustrates the schematic diagram of obtaining the wide viewing angle of the holographic image
on the basis of our deduction.
High-order diffraction beams are collected through the Fourier lens
and new hologram fringe is imaged on the image plane through one-lens imaging system [15].
For convenience, only the first-order beam along the $x$-axis is displayed.

 The complex field of the reconstructed image of the Fourier hologram is obtained from Eq. (12).
We consider the normally incident plane wave with unit amplitude.

\begin{equation}
U(x,y) =  \frac{1}{j \lambda f}
\int\!\!\!\int g_{s}(\xi,\eta)
\exp \left(-j\frac{2 \pi}{\lambda} \left[ \frac{(\xi^2 + \eta^2) \Delta z}{2f^2}
 - \frac{(x\xi + y\eta)}{f} \right] \right) d\xi d\eta
\end{equation}
where $g_{s}(\xi,\eta)$ is the sampled Fourier hologram by rectangular pixels in Fig. 5,
which has the pixel interval $p_{\xi}$ and the pixel size $\Delta p_{\xi}$ in the $\xi$-direction,
and $p_{\eta}$, $\Delta p_{\eta}$ in the $\eta$-direction.

\begin{equation}
g_{s}(\xi,\eta) = \sum_{n_{\xi}=-\infty}^{\infty} \sum_{n_{\eta}=-\infty}^{\infty}
\left[ g(n_{\xi} p_{\xi} , n_{\eta} p_{\eta} ) \textrm{rect}
\left( \frac{\xi - n_{\xi} p_{\xi}}{\Delta p_{\xi}}, \frac{\eta - n_{\eta} p_{\eta}}{\Delta p_{\eta}}
\right) \right]
\end{equation}
where rect() is a rectangular function.
Above complex field of Eq. (14) is rewritten as convolution form of two terms.

\begin{eqnarray}
U(x,y) = \frac{j}{\lambda f}
  \int\!\!\!\int g_{s}(\xi,\eta) \exp \left[ j \frac{2 \pi}{\lambda f} (x\xi + y\eta) \right] d\xi d\eta   \nonumber\\
  \ast \int\!\!\!\int \exp \left(-\frac{j 2 \pi}{\lambda} \left[ \frac{(\xi^2 + \eta^2) \Delta z}{2f^2} \right] \right)
  \exp \left[ j \frac{2 \pi}{\lambda f} (x\xi + y\eta) \right] d\xi d\eta
\end{eqnarray}

 The first line is the Fourier spectrum of the sampled hologram on the back focal plane of the lens.

\begin{eqnarray}
\int\!\!\!\int g_{s}(\xi,\eta) \exp \left[ j 2\pi \left( \frac{x \xi}{\lambda f}
   + \frac{y \eta}{\lambda f} \right) \right] d\xi d\eta = \nonumber\\
 \Delta p_{\xi} \Delta p_{\eta} \textrm{sinc} (\frac{\pi x \Delta p_{\xi}}{\lambda f})
 \textrm{sinc} (\frac{\pi y \Delta p_{\eta}}{\lambda f})
\sum_{n_{\xi}=-\infty}^{\infty} \sum_{n_{\eta}=-\infty}^{\infty}
\left[ g(n_{\xi} p_{\xi} , n_{\eta} p_{\eta} )
\exp \left\{ j \frac{2\pi}{\lambda f} ( n_{\xi} p_{\xi} x + n_{\eta} p_{\eta} y ) \right\} \right]
\end{eqnarray}
If the $g(n_{\xi} p_{\xi} , n_{\eta} p_{\eta})$ is a constant number, the summation part becomes
the Dirac comb function, which makes just the Fraunhofer diffraction pattern of the rectangular pixel array.
Inserting the value of the sampled hologram, this term turns into the Fourier spectrum image.
The summation term is represented as the Fourier transform through the Poisson summation formula.

\begin{equation}
\sum_{\alpha=-\infty}^{\infty} \sum_{\beta=-\infty}^{\infty}
G \left[ \frac{1}{\lambda f} \left( x - \frac{\lambda f}{p_\xi} \alpha \right),
\frac{1}{\lambda f} \left( y - \frac{\lambda f}{p_\eta} \beta \right) \right]
\end{equation}
From this,
Eq. (17) describes the modulation of the periodic Fourier spectrum
by the envelope of the sinc function along the $x$- and $y$-axis.
We briefly depict the modulated curve by the sinc function in Fig. 6.
If the pixel interval $p_{\xi,\eta}$ is put to be approximately equal to
the pixel size $\Delta p_{\xi,\eta}$,
the zeroth-order diffraction beam well implies the zeroth-order spectrum image.
But the high-order images are centered at the minimum region of the envelope
because the main lobe of the sinc function has twice the width of other beams.
This can be overcome by the phase-shift technology,
which appropriately controls the position of the reconstructed image to the transverse axis.

 The complex field on the focal plane is distributed within each diffracted beam
at the interval of $\frac{\lambda f}{p}$ or $\frac{\lambda f}{\Delta p}$.
Equation (18) can be expressed as another form as follows.

\begin{eqnarray}
\sum_{\alpha=-\infty}^{\infty} \sum_{\beta=-\infty}^{\infty}
\mathcal{F} \mathit{} \left\{ g(\xi,\eta)
\exp \left[ j 2\pi (\frac{\alpha}{p_{\xi}} \xi + \frac{\beta}{p_{\eta}} \eta) \right] \right\} \nonumber\\
= \sum_{\alpha=-\infty}^{\infty} \sum_{\beta=-\infty}^{\infty}
\mathcal{F} \mathit{} \left\{ g(\xi,\eta)
\exp [ j k (\xi \sin \theta_{\alpha} + \eta \sin \theta_{\beta}) ] \right\}
\end{eqnarray}
where the diffraction relation, $p_{\xi} \sin \theta_{\alpha} = \alpha \lambda$ and
$p_{\eta} \sin \theta_{\beta} = \beta \lambda$, are applied.
Substitution of this result in the convolution expression of Eq. (16) yields

\begin{eqnarray}
U(x,y) =  \sum_{\alpha=-\infty}^{\infty} \sum_{\beta=-\infty}^{\infty} \frac{1}{j \lambda f}
\int\!\!\!\int g(\xi,\eta) \exp [ j k (\xi \sin \theta_{\alpha} + \eta \sin \theta_{\beta}) ] \nonumber\\
\times \exp \left(-j\frac{2 \pi}{\lambda} \left[ \frac{(\xi^2 + \eta^2) \Delta z}{2f^2}
 - \frac{(x\xi + y\eta)}{f} \right] \right) d\xi d\eta
\end{eqnarray}
This indicates that the reconstructed wave of each diffraction beam is generated from the modulated carrier
propagating in the direction at an incidence angle $\theta_{\alpha,\beta}$ to the optical $z$-axis.
From the description in Section 2,
we find that high-order diffraction beams in this system effectively create the new viewing zone
with different perspective of the object.

 Figure 4 shows that each diffraction beam is collected on the image plane.
The focal plane of the screen lens coincides the back focal plane of the Fourier lens,
which enables the image formation as like the 4$f$ imaging system.
This one-lens imaging system can be replaced with the 4$f$ imaging system.
Thus, the hologram fringe as well as pixel pattern on the object plane is imaged on the image plane [13].

\begin{equation}
U_i(x,y) = \frac{1}{M} \left[ g\left(\frac{x}{M}, \frac{y}{M} \right) \right]
\end{equation}
where $U_{i}$ and $M$ are the field distribution on the image plane and the magnification of the image, respectively.
The magnification is determined by the ratio of distances of object and image position from the Fourier lens.
The original form of the pixel pattern can be recovered if the lens aperture is very large to use sufficient high-order beams.

 Here, the hologram fringe image of the 1st order beam dose not become the replication of their original form,
because the diffractive wave is a modulating signal by a spatial carrier, $f_{c}$.
All the diffractive waves create new hologram fringe with higher spatial frequency.
The hologram fringe makes it possible to retrieve the holographic image with different perspectives
according to the diffraction direction.
As depicted in Fig. 4,
we can observe wider viewing zone of the reconstructed image if many order diffraction beams are used.
In real system,
the spatial filtering system on the focal plane shall be required to eliminate the noise and conjugate terms
and control the beam intensity.

\section{Numerical analysis of the sampled hologram with pixel structure}
\subsection{Analysis of high-order diffraction images}
\label{NRSH}

 The outline of the sampled hologram with pixel structure for the numerical analysis is illustrated in Fig. 7(a).
The hologram is sampled in the rectangular pixel array.
For obtaining enough bandwidth of the Fourier spectrum to investigate high-order diffraction beams,
each pixel is divided into several subpixels.
This makes it possible to display high-order images in the area of the reconstruction picture.
Figure 7(b) is the Fourier transformed image of the pixel array.
As expressed in Eq. (17), the transformed image is the Dirac combo function.
When the hologram data is encoded in pixel array,
the hologram image is generated.

 The size of the pixel array is 200$\times$200 and the interval of the pixels, $p$ is set to be 4 $\mu$m.
The number of subpixels is put to be ten,
and then, the total resolution of the device encoding the sampled hologram becomes to be 2000$\times$2000.
As shown in Fig. 8,
we prepared the sampled Fourier hologram with 200$\times$200 resolution, based on in Eq. (14).
The object as `HOLO' letter placed at different depth, $\Delta z$ is used.
The hologram was synthesized at a focal length of 0.1 m
by using the recording plane wave with wavelength of 632.8 nm.
Here, the border between pixels is one subpixel with 0.4 $\mu$m thickness
and the pixel size, $\Delta p$ becomes 3.6 $\mu$m.
The hologram data is recorded in the 200$\times$200 corresponding pixel array.
That is, the hologram data within a single pixel of the device has the same value.

 Figure 9 is the numerically reconstructed image by using above sampled hologram.
For convenience, the complex modulation is considered.
We find that high-order reconstructed images generate,
whose intensity varies according to the envelope of the sinc function.
The ten high-order images up to fifth order are displayed in figure.
Figure 9(b) shows the central cross-section spectrum along the $x$-axis.
As discussed in Fig. 6,
there is a misalignment between high-order images and the envelope of the sinc function.
The mismatch of high-order images can be improved
by using the hologram made by adding the phase-shift component.
It is desirable that the zeroth order image is placed in the half main lobe of sinc function.
Here, half spatial bandwidth of the zeroth order diffraction is available.

 The board portion between pixels also causes the mismatch of the images.
The envelope of the sinc function and the image spectrum are placed with period of $\frac{\lambda z}{\Delta p}$
and $\frac{\lambda z}{p}$, respectively.
The ratio of the pixel size, $\Delta p$ to the pixel interval, $p$ determines
the number of images in one lobe of the sinc function.
We confirmed that the number of images increases with decreasing the ratio.
In above pixel structure,
since the ratio is 90\%, there appears the mismatch of up to 10\% per one order image.
For well matching, the pixel size should be designed very close to the pixel interval.

\subsection{Viewing zone of high-order images}

High-order images in Fig. 9 have the same plane picture,
which is consistent with interpretation of Section 2.
That is, diffraction beams only move the images to the transverse axis without their deformation.
In order to investigate the object view of high-order images,
the object image should be numerically reconstructed in the direction of the optical diffraction axis.
As expressed in Eq. (20),
since each high-order image is the reconstructed one by an inclined plane wave,
each order image is obtained from the coordinate transformation to the new $x_{\rm{o}}$-$y_{\rm{o}}$ optical axes [10,16],
in Fig. 10.

\begin{equation}
U_o(f_{x_o},f_{y_o}) = U_i(R[f_{x}-f_c, f_{y}, f_{z}])
\end{equation}
where

\begin{equation}
f_{z} = \sqrt{1/\lambda^2 - f_{x}^2 - f_{y}^2}
\end{equation}

 The images reconstructed in the several diffraction directions are appeared in Figure 11.
The tilted angle is determined by diffraction order, where for visualization,
large diffraction angle was adopted.
The numerical simulation is carried out for each diffraction order independently,
where the interpolation operation is applied [16,17].
We used the Fourier hologram made by two object letters with different depth, $\pm$2 mm with respect to a focal length.
The reconstructed image is shrunk in the $x_{\rm{o}}$-axis direction,
which is because the reconstructed images can be regarded as the projection images to the tilted plane.
We find that especially the object view varies according to
the diffraction order, as depicted in Fig. 3.
However, in the Fresnel hologram we confirmed that the object view is deformed as described previously.
The variation of the viewing angle shall be clarified by the numerical simulation of three-dimensional object.

 Finally, we can extend the viewing zone angle of the reconstructed image by collecting diffraction beams
to the image plane.
For this method, all of diffraction beams should have the same intensity to form the uniformly extended viewing-zone.
The spatial filter placed on the focal plane is one of methods for adjusting each beam intensity and controlling unwanted beams.
Furthermore, the spectrum intensity of a particular order can be arbitrarily controlled
by the design of the diffracting grating.
Therefore, the realization of this holographic display depends on the design of the pixel structure
sampling the hologram data.

\section{Conculsion}
\label{CU}

 The Fourier hologram makes it possible to retrieve three-dimensional image
with other perspective by a tilted plane wave.
In the sampled Fourier hologram with pixel structure,
the viewing zone angle of the reconstructed image is effectively increased
only by collecting high-order diffraction beams with different viewing direction.
The design of the pixel structure creating high-order diffraction beams
uniformly distributed is important for this holographic display.
Further study for making various types of diffraction beams is required.
This becomes one of the methods for overcoming the low spatial frequency making the narrow viewing angle.

\begin{figure}
\vspace{2.0cm}
\centerline{\epsfysize=5.53cm\epsfxsize=14.5cm\epsfbox{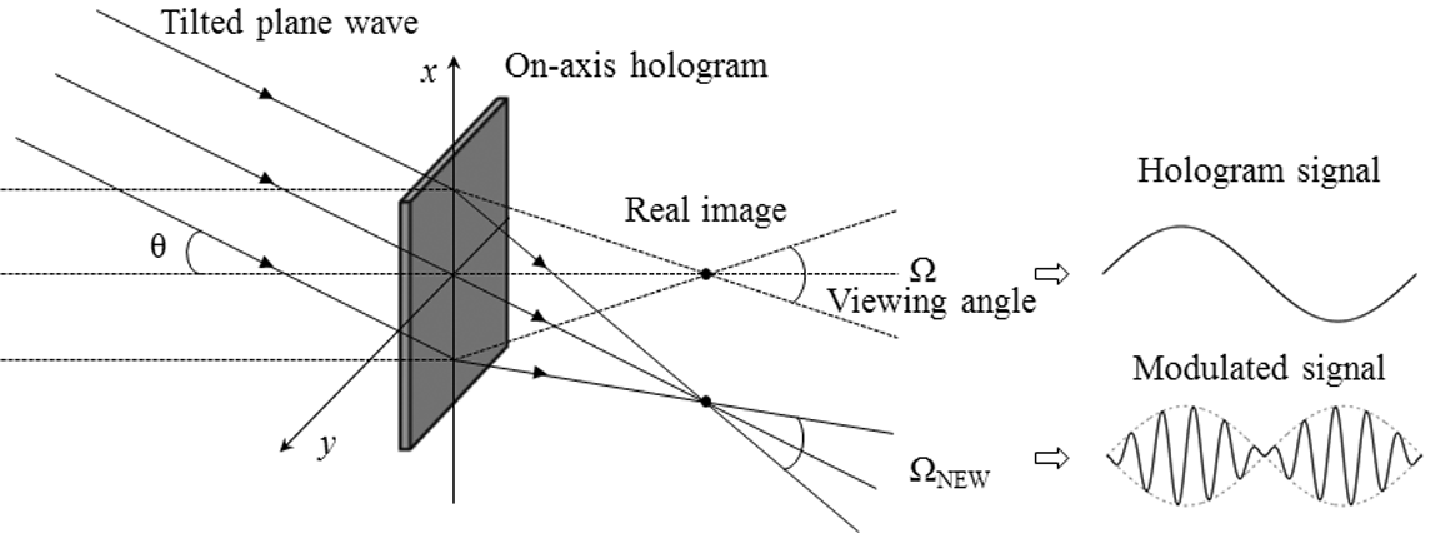}}
\vspace{1.0cm}
\caption{The reconstruction process of the on-axis hologram by a coaxial plane wave and a tilted plane wave.}
\label{f1}
\end{figure}

\begin{figure}
\vspace{2.0cm}
\centerline{\epsfysize=12cm\epsfxsize=10.69cm\epsfbox{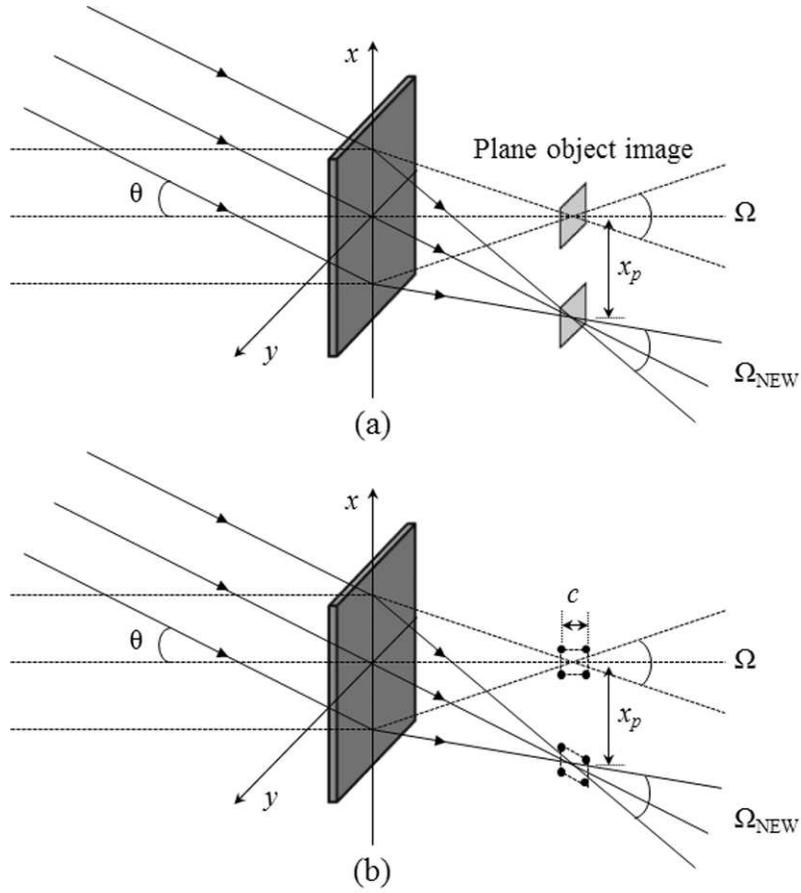}}
\vspace{0.5cm}
\caption{(a) Viewing angle change of the plane object image and (b) deformation of the image
 along depth direction reconstructed by a tilted plane wave incident on the on-axis Fresnel hologram.}
\label{f2}
\end{figure}

\begin{figure}
\vspace{0.5cm}
\centerline{\epsfysize=5.32cm\epsfxsize=11cm\epsfbox{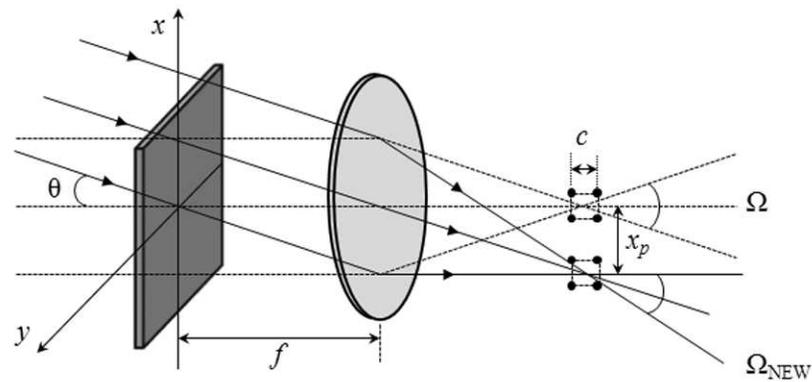}}
\vspace{1.0cm}
\caption{Viewing angle change of the object image reconstructed by a tilted plane wave incident on the on-axis
 Fourier hologram.}
\label{f3}
\end{figure}

\begin{figure}
\vspace{2.0cm}
\centerline{\epsfysize=7.35cm\epsfxsize=14cm\epsfbox{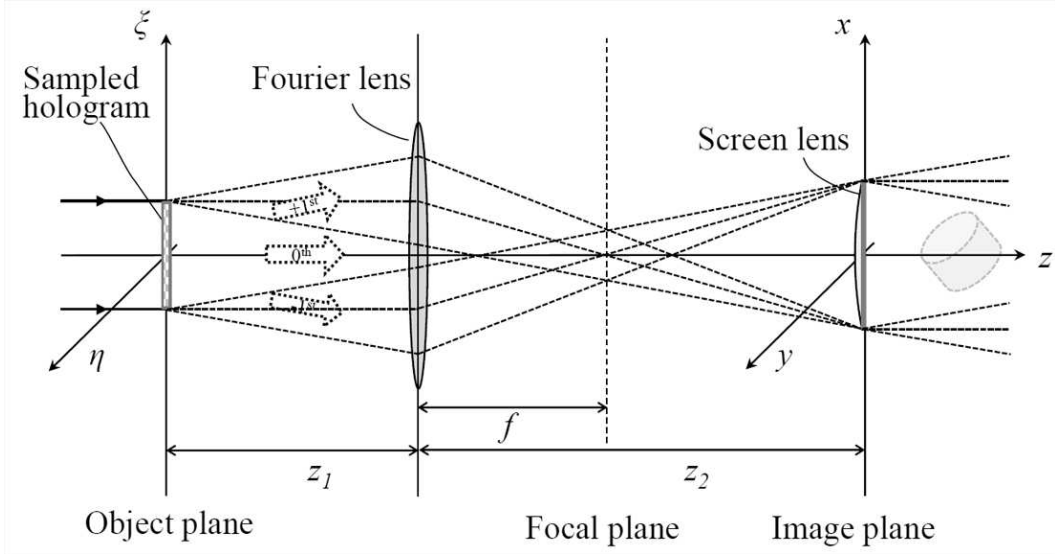}}
\vspace{1.0cm}
\caption{Schematic diagram of obtaining the wide viewing angle of the hologram image
for collecting high-order diffraction beams to the image plane.}
\label{f4}
\end{figure}

\begin{figure}
\vspace{0.0cm}
\centerline{\epsfysize=6.61cm\epsfxsize=7cm\epsfbox{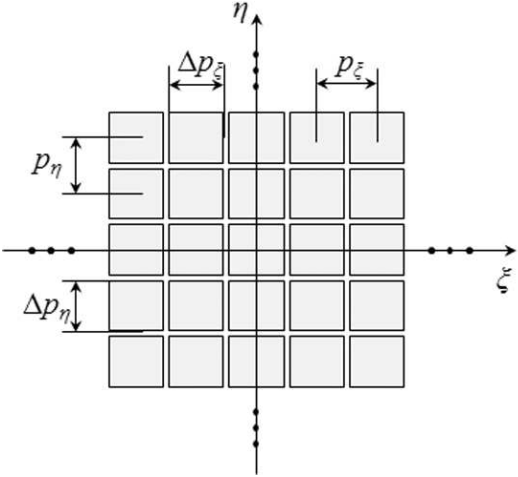}}
\vspace{0.0cm}
\caption{The pixel structure of the sampled hologram.}
\label{f5}
\end{figure}

\begin{figure}
\vspace{1.0cm}
\centerline{\epsfysize=10.49cm\epsfxsize=12cm\epsfbox{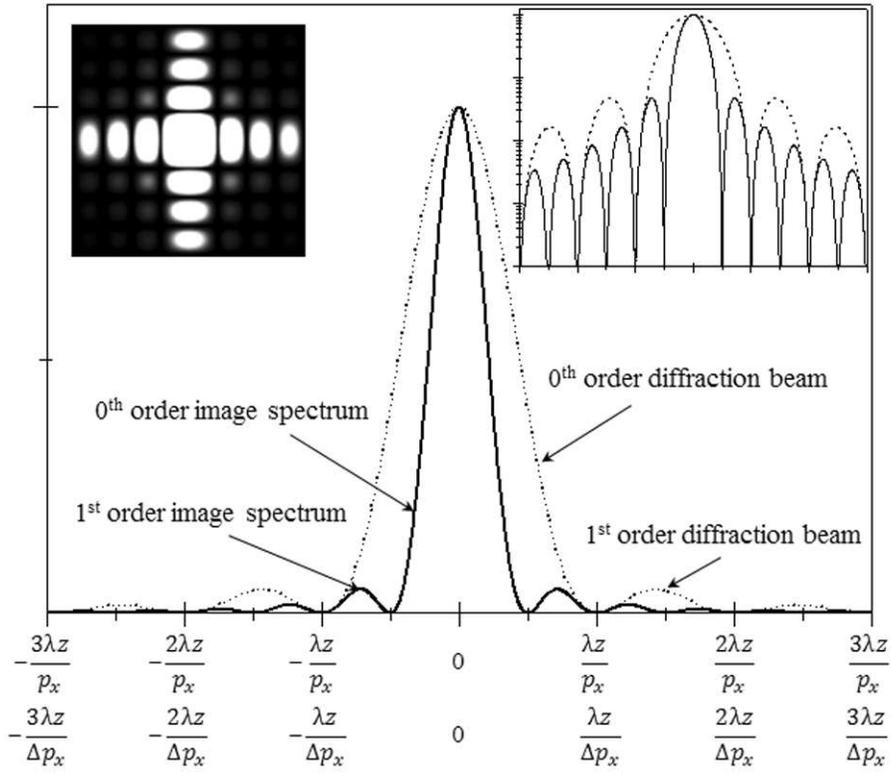}}
\vspace{1.0cm}
\caption{Modulation of the periodic Fourier spectrum by the envelope of the sinc function along the $x$-axis.
Right inset is the diffraction beam images. Left inset is the resized figure to the logarithmic scale.}
\label{f6}
\end{figure}

\begin{figure}
\vspace{2.0cm}
\centerline{\epsfysize=10.02cm\epsfxsize=9.99cm\epsfbox{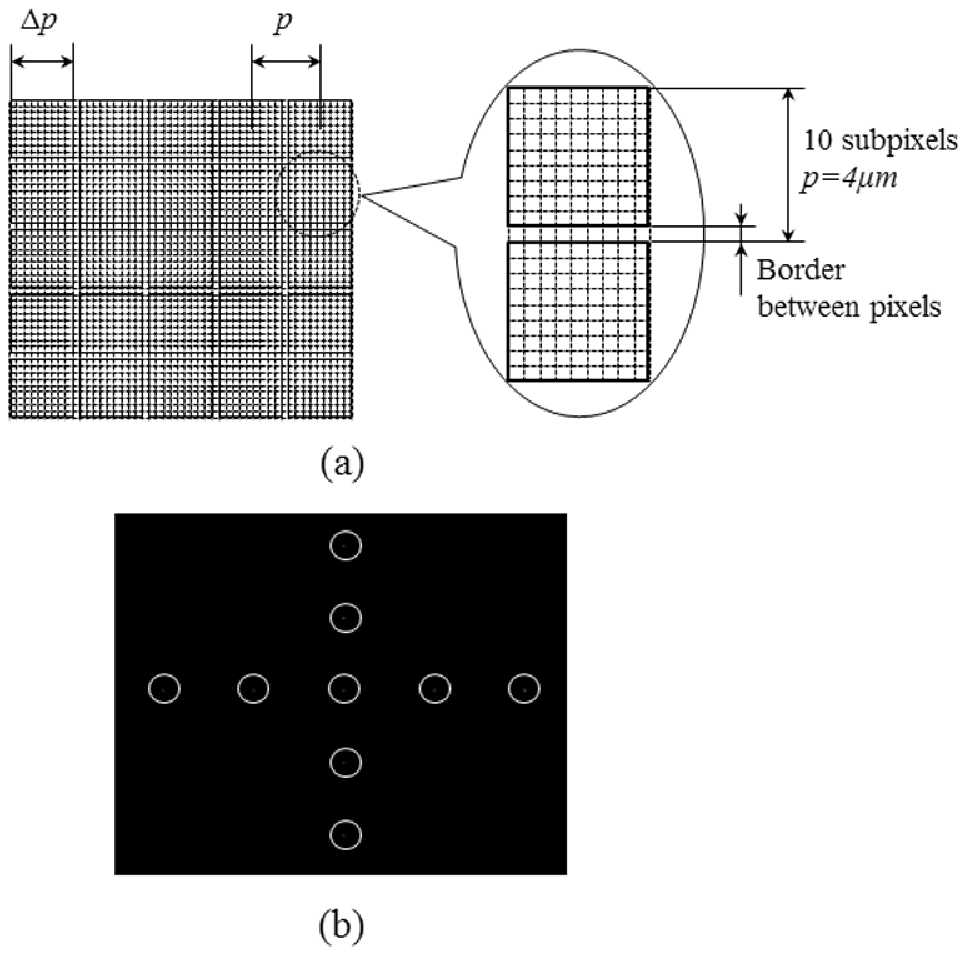}}
\vspace{1.0cm}
\caption{(a) Outline of the pixel array sampling the hologram for the numerical analysis 
and (b) the Fourier transformed image of the pixel array. The spot is located in the center of circles.}
\label{f7}
\end{figure}

\begin{figure}
\vspace{2.0cm}
\centerline{\epsfysize=4.72cm\epsfxsize=9.99cm\epsfbox{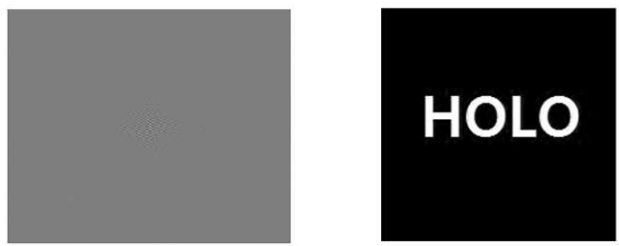}}
\vspace{1.0cm}
\caption{(a) Sampled Fourier hologram with 200$\times$200 resolution prepared from (b) the object of `HOLO' letter.}
\label{f8}
\end{figure}

\begin{figure}
\vspace{2.0cm}
\centerline{\epsfysize=12cm\epsfxsize=9.3cm\epsfbox{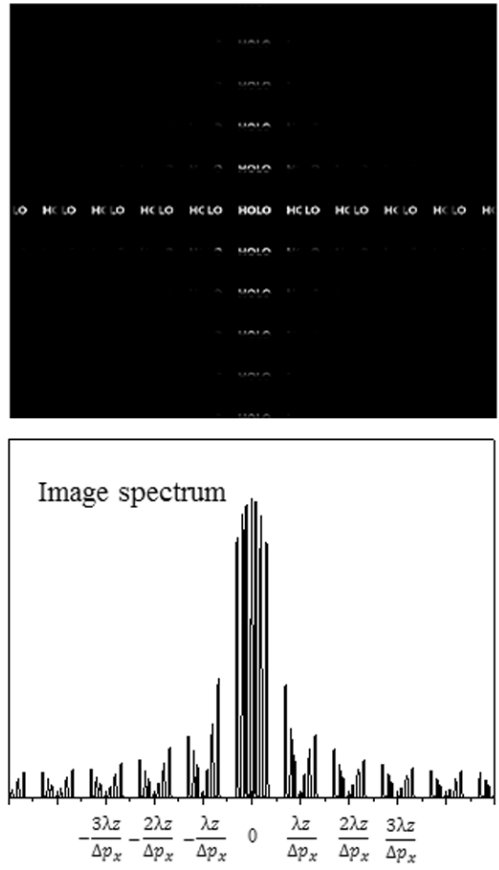}}
\vspace{1.0cm}
\caption{Numerically reconstructed images and image spectrums for the sampled Fourier hologram.
Central cross-section spectrum along the $x$-axis is presented.}
\label{f9}
\end{figure}

\begin{figure}
\vspace{5.0cm}
\centerline{\epsfysize=5.25cm\epsfxsize=9.99cm\epsfbox{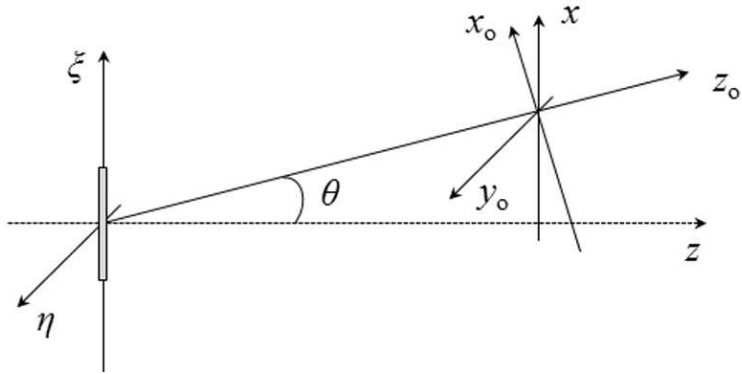}}
\vspace{1.0cm}
\caption{Geometry of the coordinate transformation to the diffraction direction.}
\label{f10}
\end{figure}

\begin{figure}
\vspace{2.0cm}
\centerline{\epsfysize=7.25cm\epsfxsize=8cm\epsfbox{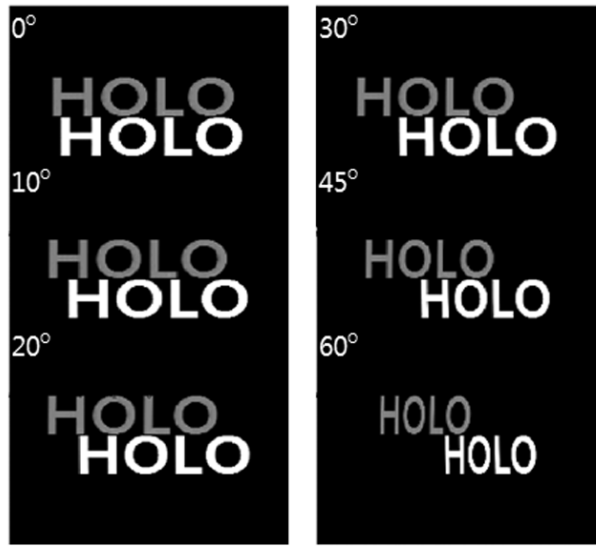}}
\vspace{1.0cm}
\caption{Viewing zones of high-order diffraction images reconstructed in the propagating direction.}
\label{f11}
\end{figure}

\end{document}